  \documentstyle[editedvolume,psfig,numreferences]{crckapb} 
  \newcommand{\msun}{M_{\odot}}
  \newcommand{\eos}{equation of state~}
  \newcommand{\eoss}{equations of state~}
  
  \newcommand{\eosp}{equation of state}

  \newcommand{\beqn}{\begin{eqnarray}}
  \newcommand{\eeqn}{\end{eqnarray}}

  \newcommand{\gpercm}{${\rm~gm/cm^3}$}
  \newcommand{\bag}{B^{1/4}}
  \newcommand{\edrip}{\epsilon_{\rm drip}}
  \newcommand{\courtesy}{~Reprinted with permission of Springer--Verlag 
  New York; copyright 1997}
  
  \newcommand{\doe}
  {This work was supported by the Director, Office of Energy Research,
  Office of High Energy and Nuclear Physics, Division of Nuclear Physics,
  of the U.S. Department of Energy under Contract DE-AC03-76SF00098.}

  \begin{opening}
  \title{INTERNAL CONSTITUTION OF NEUTRON AND STRANGE STARS}
  
  \author{NORMAN K. GLENDENNING}
  \institute{Nuclear Science Division and\\ 
  Institute for Nuclear and Particle Astrophysics\\
  Lawrence Berkeley Laboratory\\
  University of California\\
  Berkeley, California 94720}
  
  \end{opening}
  
  \runningtitle{NEUTRON AND STRANGE STARS}
  
\begin{document}
  


\begin{figure*}[tbh]
\vspace{-.3in}
\begin{center}
\leavevmode
\hspace{-.5in}
\psfig{figure=ps.lipari0,width=2.5in,height=3.in}
\end{center}
\end{figure*}

\begin{quote}
\begin{center}
{\bf TWO LECTURES \\ NATO ADVANCED STUDY INSTITUTE\\
``The Many Faces of Neutron Stars''\\Lipari, Sicily, October 1996}\\[1ex]
(Editors: M. A. Alpar, R. Buccheri, H. Ogelman and J. van Paradijs)
\end{center}
\end{quote}

\clearpage
\begin{center}
Abstract
\end{center}

\begin{quote}
In the  first of
these two lectures I will discuss the rich constitution of neutron stars
as a consequence of the Pauli principle which is engaged 
by the dominance of gravity over the nuclear force. Three especially
interesting phenomena are discussed in this contect---(1) a mechanism for
the formation of low-mass black holes distinct in their mass-range
from the black holes formed in the prompt collapse of an entire star,
(2) a multilayered crystalline structure consisting of
confined hadronic matter embedded in a background of 
deconfined quark matter (or vice versa)  which occupies
a many kilometer thick inner region, and (3) a clean and pronounced
signal of the formation of quark matter in the interior of neutron stars.
In the second lecture I will discuss the strange matter hypothesis, 
its viability as well as its consequences for compact stars and a new 
family of white dwarfs with  dense nuclear matter central regions
some orders of magnuitude greater than in ordinary white dwarfs.
\end{quote}

  \section{Neutron Stars}
  It is interesting to reflect on the consequences of the strong binding
  of the typical neutron stars that we know as pulsars. The weakest
  force -- gravity -- binds  a nucleon in a
  neutron star 10 times more strongly
   than
   the strong force binds a nucleon in  a nucleus. In doing so it works against
   the strong short-range  repulsion of the nuclear force and against
   the Fermi pressure.  Gravity therefore brings the Pauli principle into
   play  in distributing the conserved baryon number of the star over
   many baryon species so as to reach the ground state of charge neutral 
   matter.\footnote{For degenerate Fermi systems, such as 
   neutron stars and white dwarfs,
   the Fermi momentum of a given species is related to 
   the density of that species by $k\propto \rho^{1/3}$. If the corresponding
   Fermi energy lies higher than the mass of some other Fermion species
   (modified by interaction energies), it will be favorable for some Fermions
   of the first species to transform to the second (say by weak interaction).
   The same number of
   Fermions  distributed now over several species  each have
   a lower  Fermi momentum (and Fermi energy)
   than that of a single species of the same Fermion number.
   The total energy will be
   lowered as a result.}
   The name ``neutron star''
   therefore has to be understood as a generic name for a
   star populated by many baryon species, also by quarks and also
   a mixed phase of confined and deconfined matter that, as I will discuss,
   arranges itself in a very intricate pattern in the deep interior of the
   star. 
   
   I will discuss several consequences of the rich constitution of
   compact stars: (1) a mechanism for the formation of low-mass black holes
   ($M\sim 1.5-2$~M$_{\odot}$), (2) a multilayered crystalline structure of
   confined and deconfined\footnote{By deconfined phase of quarks we mean that
   quarks are asymptotically free over extended regions. This phase is also
   called the quark matter phase. By confined phase we mean the phase in which
   quarks are confined in hadrons.}
   quark phases and (3) the effect on pulsar braking indices
   of the deconfinement phase transition
   as well as the general effect of
   rotational distortion on   inferred
   magnetic fields and spin-down times of millisecond pulsars.
  									    
  In this paper we explore possibilities---not certainties.
  The properties of matter
  at densities higher than nuclear are essentially unknown, although
  they are the subjects of investigation at several ultra high energy
  accelerators.  Essentially all we can be fairly confident of are: (1) the
  \eos of dense matter obeys the condition of causality,  (2)
  the \eos also obeys the condition
  of microscopic
  stability ($dp/d\epsilon \ge 0$)
  known as Le Chatelier's principle and (3)  at  sufficiently
  high density, asymptotic freedom of quarks is achieved.
  Beyond this, a theory of
  dense matter ought to be firmly anchored to what is known at nuclear
  density. Within these constraints we explore what is allowed by the
  laws of physics, in the belief that the laws of nature are likely to
  be realized in many if not all possible ways in the Universe.
  The vehicle for the exploration is a covariant nuclear field theory
  that embraces the above constraints \cite{glen85:b,book}.
  
  Gravity compresses matter severely in a star of mass
  typical of those that are known, say nominally 1.4~M$_{\odot}$
  Although the strong force resists the compression, matter at the densities
  typical of the center of a canonical neutron star has a compression energy of
  several hundred MeV per baryon. Nevertheless, the binding energy per
  baryon in a canonical star is 100 MeV (the gravitational
  less the compression energy) and it falls to smaller and even
  negative values with decreasing mass. These numbers can be read from
  Figs.\ \ref{bam} and \ref{bvsm}.
  \begin{figure}[tbh]
  \begin{center}
  \leavevmode
  \centerline{ \hbox{
  \psfig{figure=ps.lipari1,width=2.38in,height=2.76in}
  \hspace{.2in}
  \psfig{figure=ps.lipari2,width=2.38in,height=2.76in}
  }}
  \begin{flushright}
  \parbox[t]{2.25in} { \caption { \label{bam} Compression energy of
  neutron star matter.
  }} \ \hspace{.2in} \
  \parbox[t]{2.25in} { \caption { \label{bvsm} Binding energy per baryon in
  a sequence of neutron stars. Dashed line represent configurations beyond
  the mass limit. Note negative binding for low-mass stars
  \protect\cite{book}.\protect\courtesy.
  }}
  \end{flushright}
  \end{center}
  \end{figure}
  I used to think that the relatively small range in which neutron star
  masses fell had to do with the creation mechanism, the evolution of massive
  stars leading to core collapse. And perhaps it does. But inasmuch as
  a supernova is powered by the transfer of a small fraction of the
  energy carried by neutrinos
  which  themselves
  derive their energy
  from the gravitational
  binding of the neutron star, it is clear that the binding energy is
  sufficient  to power a supernova
  only for a narrow range of masses. Theoretically
  a neutron star of mass as low as
   1/10~M$_{\odot}$ could exist, but evidently a neutron star of less than about
   1/2~M$_{\odot}$ could
  not be made in the typical way.
  
  \begin{figure}[tbh]
  \begin{center}
  \leavevmode
  \centerline{ \hbox{
  \psfig{figure=ps.lipari3,width=2.38in,height=2.76in}
  \hspace{.2in}
  \psfig{figure=ps.lipari4,width=2.38in,height=2.76in}
  }}
  \begin{flushright}
  \parbox[t]{2.25in} { \caption { \label{pops_240}  Neutron star at the mass
  limit that is composed of
  hyperonized matter  (charge neutral and in general beta equilibrium).
  }} \ \hspace{.2in} \
  \parbox[t]{2.25in} { \caption { \label{pops_k240b180} Neutron star at the
  mass limit
  for which  at low density near the edge a pure
  hadronic phase exists, interior to which 
  a mixed phase of baryons and quarks exists at intermediate density,
  and pure quark matter at high density in the inner region of 4.5 km.
  }}
  \end{flushright}
  \end{center}
  \end{figure}
  The rich  hyperonized
  composition of charge neutral equilibrated neutron star matter is
  illustrated in  Fig.\ \ref{pops_240}.
  For charge neutral matter for which
  a phase transition to the coexistence phase and pure phase of quark matter
  occurs, the composition is shown
  in Fig.\ \ref{pops_k240b180}. Notice the mixed phase region
  between 4.5 and 7.8 km in which both hadronic matter and quark matter
  are in equilibrium. I will discuss later the geometrical structure that
  develops in this region.
  In each figure the particle populations are shown as a function
  of radius for the limiting mass star. In the first example, the phase
  transition to quark matter does not occur because of the choice of
  parameters, which is after all uncertain. The two cases therefore embrace
  two possibilities that may be realized in nature.
  The theory underlying the calculations can be found in Refs.
  \cite{glen85:b,glen91:c,glen91:a,book}.

  \subsection{The First Ten Seconds}
  
  The protoneutron star formed in core collapse is hot and very lepton rich
  and the matter of which it is made is far from its final composition---that
  of the ground state of cold dense matter. As the star cools and shrinks
  further during deleptonization, the Fermi energies of neutrons and protons
  rise with increasing density ($k\propto \rho^{1/3}$). It becomes
  increasingly
  favorable with increasing density
  for neutrons and protons to transform to other
  baryon number carrying species,  hyperons or quarks,
  through the weak interaction ($\tau_{{\rm weak}}\sim
  10^{-10}$ s) \cite{glen85:b,book}. 
The Fermi pressure is relieved as a result, the \eos
  is softened, and the matter of the star is less able to resist the
  pull of gravity. If the collapsing core lies in mass above the limiting mass
  of the final ground state  of cold neutron star matter
  but below the limiting mass of the hot lepton
  rich protoneutron star, it will continue to collapse to form a low-mass
  black hole \cite{glen94:d}. 
But its continued collapse progresses on the time scale of
  neutrino diffusion (10 s) and is conditioned by  neutrino loss.
  Therefore, unlike the prompt collapse of the entire star to form
  a black hole of several tens of solar mass, these low-mass black holes
  are formed after the supernova explosion and neutrino pulse. Depending
  on the mass function of massive stars and the dependence of core mass
  on stellar mass, a large fraction of massive stars may end their lives in
  a supernova explosion and a residual low-mass black hole instead of a
  neutron star.

  To estimate the mass or baryon number window for such events, we
  may compare neutron star sequences for equilibrated n,p,e matter and
  fully equilibrated matter corresponding either to hyperonized matter, or
  to  partially or wholly deconfined quark matter.
  I refer to  stars of the latter
  type as hybrid stars because they have an ordinary neutron star
  exterior, an intermediate region of mixed quark and hadronic matter,
  and possibly a pure quark matter core
\cite{glen91:d,glen91:a,glen95:c}.
  Stellar mass as a function of density is shown in Fig.\ \ref{mass_3}.
  The three stellar sequences illustrated correspond to different degrees
  of completion of equilibrium.
  The softening of the equation of state due to hyperonization is quite apparent.
  (It is also apparent that models of neutron stars which neglect complete
  equilibrium are unrealistic.)
  
  Figure \ref{mas_col_240} illustrates stellar
  mass as a function of baryon number  for a protoneutron star
  (here modeled as a n,p,e star) and a fully equilibrated configuration
  corresponding to a hyperonized star. The window between H and P  corresponds
  to the mass or baryon number of protostars that will collapse following
  a supernova and neutrino display to a low-mass black hole.
  \begin{figure}[htb]
  \begin{center}
  \leavevmode
  \centerline{ \hbox{
  \psfig{figure=ps.lipari5,width=2.38in,height=2.76in}
  \hspace{.2in}
  \psfig{figure=ps.lipari6,width=2.38in,height=2.76in}
  }}
  \begin{flushright}
  \parbox[t]{2.25in} { \caption { \label{mass_3} Three stellar sequences
  as a function of central density. Three stages of completeness with respect to
  beta equilibrium are illustrated \protect\cite{book}.\protect\courtesy.
  }} \ \hspace{.2in} \
  \parbox[t]{2.25in} { \caption { \label{mas_col_240} Two stellar sequences
  as a function of baryon number. The one extending to P represents a protostar,
  and the one to H a fully equilibrated neutron star in its ground state.
  A core collapse with A falling between H and P will form a low-mass
  black hole \protect\cite{glen94:d}.
  }}
  \end{flushright}
  \end{center}
  \end{figure}
  
  If the star deleptonizes to a stable mass it will have cooled to the
  MeV level or less and will essentially be frozen for eternity as far as
  nuclear transformations are concerned. Those stars that lie close to the mass
  limit will have a rich baryon population, either hyperons or quarks or both as
  illustrated in Figs\ \ref{pops_240} and \ref{pops_k240b180}.
  The neutrino display accompanying neutron star formation and low-mass
  black-hole formation will differ in the tale of the signal, the
  neutrinos suffering an extreme redshift  in the latter case.
  
  For reasons not well understood, neutron stars have a high average velocity
  of about 500 km/s \cite{lyne94:a}. 
  So would the low-mass black hole formed in the
  prompt collapse of the protoneutron star. It is interesting to contemplate
  possible differences in the interaction with the interstellar medium.
  A neutron stars  produces a bow shock fanning out to hyperbolic
  wings whose visible presence is revealed by the H alpha line.
  What shock pattern would a high-velocity black hole produce?

  \subsection{Crystalline Structure in Stars}
  
  I turn now to other possible consequences of the high compression of matter
  in neutron stars---the formation of 
  an unusual crystalline region consisting of
  confined nuclear matter and deconfined quark matter.
  For neighboring mass stars the regions of varying crystalline
  structure extending over  a radial distance of many kilometers
  is illustrated in Figs. \ref{pie} and \ref{pie2}.
  I describe the situation below.
  \begin{figure}[tbh]
  \begin{center}
  \leavevmode
  \centerline{ \hbox{
  \psfig{figure=ps.lipari7,width=2.38in,height=2.76in}
  \hspace{.2in}
  \psfig{figure=ps.lipari8,width=2.38in,height=2.76in}
  }}
  \begin{flushright}
  \parbox[t]{2.25in} { \caption { \label{pie} Pie section of a hybrid
  star showing regions of quantum liquid (white areas) and solid regions
  of various geometric phases \protect\cite{book}.\protect\courtesy.
  }} \ \hspace{.2in} \
  \parbox[t]{2.25in} { \caption { \label{pie2}  Similar to \protect\ref{pie}
  but for a slightly less massive star \protect\cite{book}.\protect\courtesy.
  }}
  \end{flushright}
  \end{center}
  \end{figure}
  
  A possible phase transition from quarks confined in hadrons to
  deconfined quark matter in which the quarks are essentially free to
  move in an extended colorless region was discussed by many authors
  beginning in the mid seventies
  \cite{baym76:a,chapline76:b,keister76:a}
  and right up to the present. However
  a profound change has taken place in the understanding of the nature of the
  phase transition as a result of work that I published in 1991-1992
\cite{glen91:a}.
  Originally, it was imagined that the phase transition was a constant pressure
  one like the conversion of water into steam.  In such phase transitions,
  the nature of the two
  phases remains unchanged until the transition is complete
  from one pure phase to the other.
  But such phase transitions
  are a very special case of one-component substances.

  Neutron stars are not made from a single-component substance.
  There are, in fact, two independent components or conserved quantities
  that characterize the matter of a star---the original 
  baryon number and its net charge.
  A star has zero {\sl net} charge because above an infinitesimal
  ratio of net charge to baryon number the Coulomb force would repel
  additional charged particles  (the Coulomb force being so much stronger than
  the gravitational). Note that neutrality is a global condition, not a local one.
  The mistake of enforcing charge neutrality in stellar models
  as a local constraint  has been made time and again. It is not the
  charge density $q(r)$ that must vanish but only $\int q(r) r^2 dr$
  that must vanish. The latter is a less restrictive condition, and if the
  internal forces can take advantage of the freedom admitted by global
  neutrality to achieve a lower energy state, the internal forces will do so.
  A well known example is an atom which is neutral but has finite
  charge density. 
  
  The appropriate way to express global neutrality of two
  {\sl uniform} 
  substances in contact and in equilibrium with each other, such as quark matter
  and confined nuclear matter,  is
  \beqn
  4\pi\int_V  q(r) r^2\, dr =
  (V-V_Q)  q_H (\mu_b, \mu_q) + V_Q q_Q(\mu_b, \mu_q) =0\,.
  \label{Q}
  \eeqn
  Because the substances are uniform in any small locally inertial region
  $V$ of the
  star,  the integral over densities that expresses global neutrality
  takes this simple form. The baryon and electric charge chemical potentials
  which characterize the state of the phases are the arguments
  and the $V$'s denote volumes. 
  Quark chemical potentials are related to the baryon and 
  charge chemical potentials in the usual way
   \beqn
    \mu_u =\mu_c = (\mu_b - 2\mu_q)/3,~~~~~\mu_d = \mu_s =
     (\mu_b +\mu_q)/3\,.
      \eeqn
  Similarly to (\ref{Q}) the expression for overall baryon conservation
  within (an unknown) volume $V$ containing $B$ baryons
  is
  \beqn
  (V-V_Q)  \rho_H (\mu_b, \mu_q) + V_Q \rho_Q(\mu_b, \mu_q) =B
  \label{B}
  \eeqn
  where $\rho$ denotes baryon number density.
  The Gibbs condition for equilibrium is
  \beqn
  p_H(\mu_b,\mu_q,T)=p_Q(\mu_b,\mu_q,T)
  \label{pres}
  \eeqn
  We have here three equations (\ref{Q}, \ref{B}, \ref{pres}) in the
  unknowns $\mu_b,~\mu_q$ and $V$ for any chosen proportion
  \beqn
  \chi \equiv V_Q/V
  \eeqn
  of quark phase. Since $\chi$ appears explicitly in the equations that
  define the solution ($V_Q=V\chi$),
  the solution changes as the proportion and since the
  chemical potentials determine the state of quark and hadronic matter,
  the properties of the
  phases change with proportion. In particular, the concentrations
  of electric charge to baryon number changes in each phase as the proportion
  of quark matter changes. That is
  \beqn
  c_H\equiv  \frac{q_H (\mu_b, \mu_q)}{ \rho_H (\mu_b, \mu_q)}\,,~~~~~
  c_Q\equiv \frac{q_Q(\mu_b, \mu_q) }{ \rho_Q(\mu_b, \mu_q)}
  \eeqn
  depend on $\chi$ through the dependence of the chemical potentials
  on $\chi$.
  [Although the concentrations of  charge to baryon number vary
  in each phase  with proportion of phases,
  overall conservation of charge and
  baryon number is guarantied by
  (\ref{Q}) and (\ref{B}).]

  Interesting and possibly far reaching consequences
  for pulsars follow from the above analysis.
  Let us enquire as to the nature of  the  force  that drives
  the system to   optimize  the
  charge concentration in the two equilibrium phases.
  Neutron matter is highly isospin asymmetric. (The proton to neutron ratio
  is far from unity.) There are two agents that will tend to drive the
  system to greater symmetry. One is the Fermi energy---any
  inequality of the Fermi surfaces
  of neutrons and protons
  corresponds to a higher energy state than one
  for which the Fermi surfaces are equal (symmetry).
  The other agent is the specific preference of the strong interaction
  for symmetry, namely the coupling of the
  rho meson to the isospin current of the nucleons.
  The valley of beta stability, well known in nuclear physics,
  attests to the preference for symmetry.
  
  On the other hand, charge
  neutrality is an overriding condition  because gravity overwhelms all other
  forces in a star.
  Consequently as long as
  neutron star matter is in the confined phase, it must be highly
  asymmetric. However, in those inner regions of the star where the pressure or
  density is high enough, some of the nuclear matter will condense into
  quark matter in equilibrium with it. The strong isospin asymmetry
  of neutron star matter
  can then be relieved by transferring charge and strangeness (mediated by the
  weak interactions) between the
  two phases in such amount as minimizes the energy.
  Of course, in chemical thermodynamics it is not necessary
  to enumerate individual reactions nor calculate rates.
  \begin{figure}[htb]
  \begin{center}
  \leavevmode
  \psfig{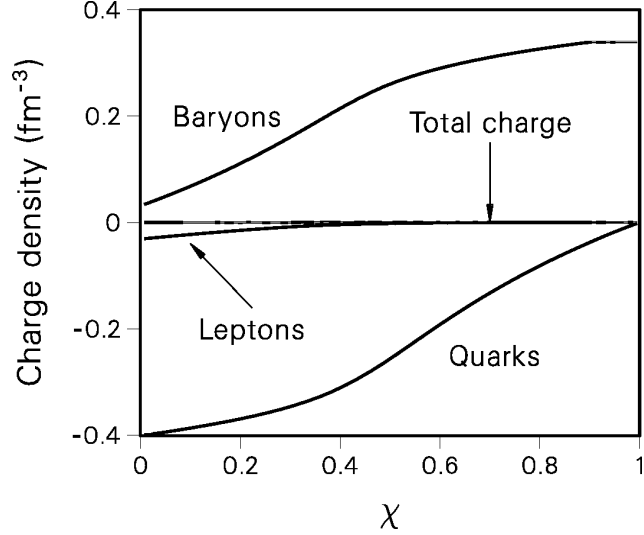}
  \parbox[t]{4.6 in} { \caption { \label{chiq} Charge carried on regions
  of quark and hadronic matter in equilibrium as a function of the proportion
  of quark matter \protect\cite{glen95:c}.
  }}
  \end{center}
  \end{figure}
                                                                                  
  For the reason discussed above,
  regions of nuclear matter will have positive charge and
  regions of quark matter, negative. The Coulomb force will tend to break up
  regions of like charge into smaller ones intermixed with regions of
  opposite charge. The surface energy will resist the breakup. The competition
  will be resolved when the rarer phase takes on a crystalline order 
  within the dominant phase. The nature of the crystalline form, its size
  and spacing, will vary as the proportion of the phases because the properties
  of each phase varies. This was proven above. Therefore the crystal 
  characteristics will vary in the changing pressure environment of a star
  and therefore as a function of position in the star.

  I have verbally described a situation which can and has been
  defined quite precisely in terms of specific models of nuclear and quark matter.
  Let us see in a schematic fashion how this can be done 
  (see \cite{glen95:c,book})
  for details).
  The surface energy per unit volume of a quark drop of radius $r$
  in a nuclear background of radius $R$, chosen so that there is zero net charge
  in $R$ (Wigner-Seitz cell) is
  \beqn
  E_S/V = [4\pi r^2 \sigma]/[(4\pi/3)R^3] = (3\sigma \chi)/r
  \equiv S(\chi)/r\,,
  \eeqn
  where for droplets, $\chi=(r/R)^3$ is the volume proportion of the quark phase.
  Likewise, while more involved to prove \cite{ravenhall83:a}, the
  Coulomb energy per unit volume has the form,
  \beqn
  E_C/V = C(\chi)r^2\,.
  \eeqn
  Their sum is a minimum  when
  the size $r$ of the droplets is such that
  $E_S=2 E_C$\,.
  The  above equations lead at once to
  \beqn
  r = \biggl( \frac{S(\chi)}{2 C(\chi)} \biggr)^{1/3},~~~~~
  R=\frac{r}{\chi^{1/3}}\,.
  \eeqn
  Thus
  at each proportion $\chi$ a definite size of quark drops immersed in the
  nuclear matter  and their spacing is specified.
  We note that the long-range of the Coulomb force is screened by the
  formation of the lattice.
  
  The functions $C$ and $S$ and the proportion $\chi$, expressed in terms
  of the geometry of the one phase immersed in the other, have quite
  definite forms  for each geometry,
  droplets,  rods, and
  slabs. Also the relationship between $\chi$ and the dimensions characterizing
  the geometry are quite definite.
  This is all quite
   analogous to  the sub-nuclear crystal structure of nuclei
  immersed in an electron gas, which is believed to form the crust
  of a neutron star \cite{ravenhall83:a} and it is somewhat surprising that
  it took so many years before it
  was realized that the mixed phase of neutron star matter in equilibrium with 
  quark matter 
  would also from a crystalline lattice.
  					     
\begin{figure}[tbh]
  \begin{center}
  \leavevmode
  \psfig{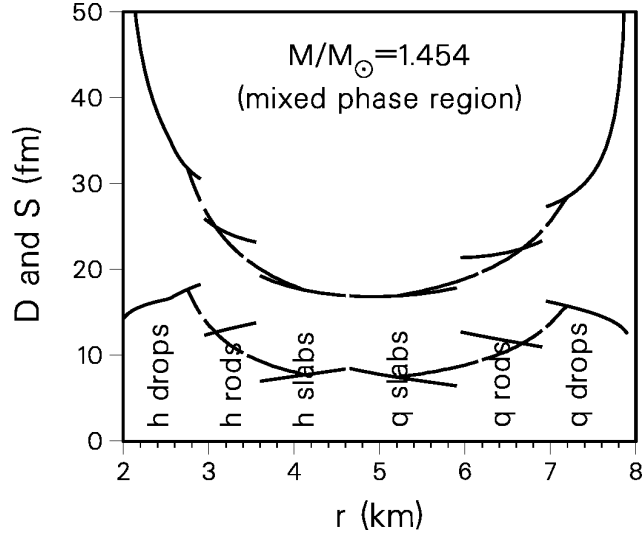}
  \parbox[t]{4.6 in} { \caption { \label{cry1} Diameter (D) and spacing (S)
  of the six idealized geometrical crystalline phases in a hybrid star.
  Note the suppressed origin. The central core is pure quark matter.
  (Dashed line is a continuous dimensionality interpolation.)
  }}
  \end{center}
  \end{figure}

  For a particular choice of nuclear properties within the range defined by
  experiment, and a particular model of nuclear and quark matter we can see
  in  Fig.\ \ref{cry1} how the diameters, spacings and geometry
  change.
  Of course the discrete geometries are only
  idealizations which are interpolated by nature. 
  The thickness and 
  locations of the geometrical phases depend very sensitively
  on the mass of the star (Figs.\ \ref{test240}, \ref{test240s}). 
  This is so because the density distribution
  in a star of canonical mass is very flat in the central region. Therefore
  a small change in mass corresponding to a small change in central density,
  implies a large radial displacement at which a  given density is to be found,
  say  the density corresponding to the boundary
  between pure quark and mixed phase.
  
  The sensitivity of the thickness and location of the crystalline structure
  is perhaps interesting in connection with pulsar glitches.
  Glitches imply the existence of solid regions in a star because a purely
  liquid or gaseous star has no means by which its moment of inertia can
  do anything but follow smoothly the change in rotational frequency as the
  star spins down due to energy losses that are generally presumed to be
  of magnetic-dipolar form.
  We believe that
  there is a thin ionic crust on neutron stars which either cracks
  from time to time or from which pinned superfluid vortex
  lines slip  catastrophically to new sites. 
  
  Now we have reason to believe that in addition to the surface crust
  there is also a many kilometer 
  thick crystalline region in the interior to which
  vortex lines could also be pinned. So a vortex line instead of being pinned
  on  the crust at opposite sides of the star could be pinned, one end at the
  crust, the other in the crystalline core.
  The great sensitivity of the geometrical nature
  and thickness of the interior solid  would give great individuality
  to the behavior of different pulsars even of very close mass. One can imagine
  phenomena involving the sympathetic
  response of one region to the other. However, I think that it will be
  very difficult to arrive
  at semi-quantitative predictions, but perhaps not hopeless.
  \begin{figure}[tbh]
  \begin{center}
  \leavevmode
  \centerline{ \hbox{
  \psfig{figure=ps.lipari11,width=2.38in,height=2.76in}
  \hspace{.2in}
  \psfig{figure=ps.lipari12,width=2.38in,height=2.76in}
  }}
  \begin{flushright}
  \parbox[t]{2.25in} { \caption { \label{test240} The radial location of
  boundaries between different phases is shown for stars of different
  mass. The dotted region is enlarged in Fig.\
  \protect\ref{test240s} \protect\cite{book}.\protect\courtesy.
  }} \ \hspace{.2in} \
  \parbox[t]{2.25in} { \caption { \label{test240s} Detail of the dotted
  region of Fig.\ \protect\ref{test240} \protect\cite{book}.\protect\courtesy.
  }}
  \end{flushright}
  \vspace{.5in}
  \end{center}
  \end{figure}
  
  \subsubsection{Evolution of Internal Structure with Pulsar Spin-Down}
  
  Another aspect of the deconfinement phase transition is interesting beside
  the statics discussed above. I refer
  to  the response of the internal structure
  of the star to changing rotational frequency (see footnote \ref{rotation}).
  A signal
  of the changing structure may show up in the braking index of pulsars. We take 
  up that subject  in the next section. 
  
\begin{figure}[tbh]
  \begin{center}
  \leavevmode
  \psfig{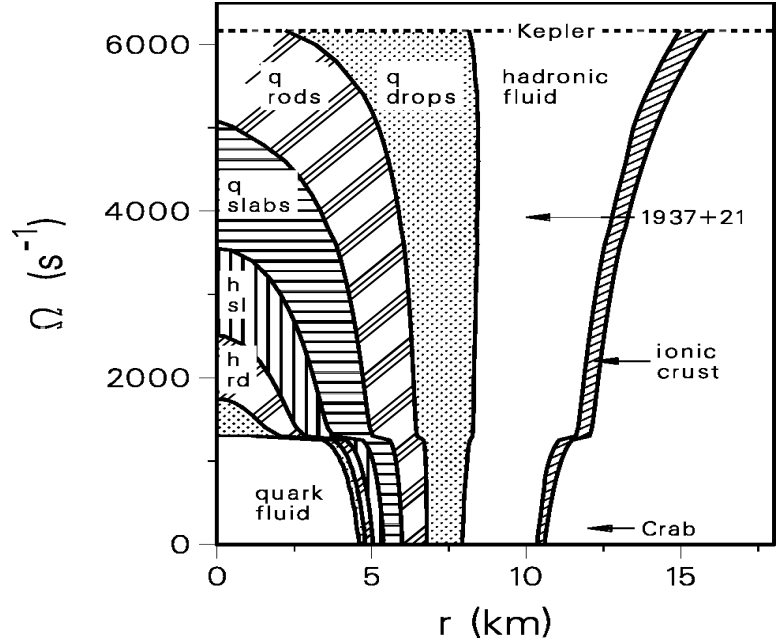}
  \parbox[t]{4.6 in} { \caption { \label{OKPOEQ_1_42_G3_B18} Evolution  of
  radial location between boundaries of various phases in a neutron star
  of given baryon number as a function of rotational
  angular velocity.\protect\footnotemark[3]
  }}
  \end{center}
  \end{figure}

  At the higher frequencies characteristic of a particular pulsar
  in its early life, the central density is suppressed compared to what
  it will become at lower frequency in later life. Imagine a combination of 
  mass and frequency such that in early life the central density is below
  the phase transition, but rises above it in later life at a lower
  frequency.
  The  radial location
  of boundaries between pure and mixed phases and between geometrical
  phases will accordingly change with angular velocity $\Omega$. 
  This is shown in Fig.\ \ref{OKPOEQ_1_42_G3_B18}.  For the particular
  stellar model and stellar mass, the central density rises to the
  transition density to pure quark matter at the center of the star
  at angular velocity of about $\Omega=1250$~rad~s$^{-1}$.
  Because of the flat profile of neutron stars near their center,
  the phase boundary moves outward to larger radius with small change in 
  angular velocity. The radius of the star meanwhile shrinks.

  \subsection{Braking Index and Internal Structure~$^3$~}
  \footnotetext[3]{From 
  unpublished
  work of N. K. Glendenning, S. Pei and F. Weber
\cite{glent97:a,glent97:b}.}
  
  Pulsar slow down is usually represented by an energy loss equation of the
  form
  \beqn
   \frac{dE}{dt} =
    \frac{d}{dt}(\frac{1}{2} I \Omega^2 ) =
     - C \Omega^{n+1}
      \label{energyloss}
       \eeqn
        where, for magnetic dipole radiation,
         $ C=  \frac{2}{3} m^2 \sin^2 \alpha $,
         $n=3$ ,
          $m$ is the magnetic dipole moment and $\alpha$ is the angle
          of inclination between magnetic moment and rotation axis.
          We shall refer to $n$ appearing in the energy-loss equation as the
          {\sl intrinsic} index.
  Other multipoles may participate but it can be expected that one will dominate
  (magnitude of $C$). Other variables  may play a role in the
  radiation from a pulsar over its lifetime but the response of the
  moment of inertia to the changing rate of rotation will produce its own
  effect, upon which other variables will superpose theirs.

  Usually the above equation is represented by
  \beqn
  \dot{\Omega}=-K \Omega^n
  \label{braking}
  \eeqn
which follows from
(\ref{braking}) if $I$
is a constant independent of frequency and therefore of time and where
$K=C/I$  and $n$ is   the braking index.
  From these equations
   follow the well known pulsar spin-down time (or age) and
   estimate of the magnetic field strength.
   However, the moment of inertia of a rotating
  star depends on its frequency and therefore on time. The dependence
  is very complicated. In General Relativity the very metric of spacetime
  is affected by the rotation of a star: local inertial frames are set into
  rotation. The centrifugal effects on the star are measured with respect to
  the angular frequency of the local frames. These depend on distance
  from the center of the star. Consequently, not only
  is a star's shape flattened as it would be in classical physics,
  but its internal structure is altered -- the distribution of energy density
  and hence of all other constituents of the star -- the location of the
  thresholds for various baryon species and the boundaries of different 
  phases.\setcounter{footnote}{3}\footnote{\label{rotation}The usual 
  expression for the moment of inertia
  in General Relativity is not adequate for our purpose. It ignores the
  dragging of local inertial frames, the alteration of the metric by rotation,
  and even the centrifugal flattening. Instead we must use an expression
  that incorporates these effects as derived by Glendenning and Weber
  \cite{glen92:b,glen93:a}.} 
  
  If the frequency dependence, and hence time dependence, of the moment of inertia
  is taken into
  account, as it should, especially for rapidly
  rotating pulsars, 
  the rate of change of angular velocity (\ref{braking}) is replaced by
  \beqn
  \dot{\Omega}= -K(\Omega) \Omega^n \Bigl(1  + \frac{I^{\prime}(\Omega)\,
  \Omega}{2I(\Omega)}
   \Bigr)^{-1}
     \label{braking2}
     \eeqn
 where $K$ is no longer a constant because of the
    angular velocity dependance of $I$ and
       $I^\prime \equiv dI/d\Omega$.
  
      Equation (\ref{braking2})
  explicitly shows
   that the angular velocity dependence of  $\dot{\Omega}$
   corresponding to {\sl any}
   mechanism that absorbs (or deposits) rotational
   energy such as
   \ (\ref{energyloss})
   cannot be a power law, as in  (\ref{braking}) with
   $K$ a constant. It must depend on the
   mass and internal constitution of the star through the
   response of the moment of inertia to rotation. Intuitively it is clear that
   over the era of observation, and even much longer, the moment of inertia
   is essentially a constant. 
This does not alter the fact that the law governing
   the decay of the angular velocity
 is (\ref{braking2}) and not (\ref{braking})
   because $I^\prime$,  which is also constant over any observational era,
   is nonetheless {\sl finite}.
  We shall see that  the magnitude of $I^\prime \Omega / (2I)$ will 
  affect  the rate of pulsar spin down  differently
  in different eras of a its
  life because the  internal constitution of the star  changes
  with changes in the  density or pressure profile
  caused by the centrifugal force.

The dimensionless measurable quantity $\Omega \ddot{\Omega} /
\dot{\Omega}^2$ equals the intrinsic
index of the energy loss mechanism only for
$I=$ constant, or for $\Omega \rightarrow 0$ as can be found from
(\ref{braking}). Otherwise we find from
(\ref{braking2}) that
   \beqn
   n(\Omega) \equiv\frac{\Omega \ddot{\Omega} }{\dot{\Omega}^2}
  = n
   - \frac{ 3  I^\prime \Omega +I^{\prime \prime} \Omega^2 }
   {2I + I^\prime \Omega} \,.
   \label{index}
   \eeqn
    The measurable braking index $n(\Omega)$ can be very different  from $n$;
 it can even be zero or have negative values depending on the {\sl
 derivatives} of
 the moment of inertia and on $\Omega$.
 If $I^\prime \Omega$
 dominates the other terms in (\ref{index}), then $n(\Omega)$ will vanish:
 If $I^{\prime\prime} \Omega^2 > 6I$ then the observable braking index
 will be negative for magnetic dipole radiation. 

 Because the  braking index  (\ref{index})  depends explicitly and
 implicitly on $\Omega$, even if the energy loss mechanism remained
 unchanged during the entire life of a pulsar, its measured
 index (\ref{index}) will change with time, in general continuously, but under
 circumstances that we discuss later, it can change radically over
   an era in the life of the star.
   The right side  of (\ref{index}) reduces to
   a constant $n$ only if $\Omega=0$ or $I$ is independent of angular 
velocity. But his cannot
   be, except for slow pulsars.
   The centrifugal force insures the response of $I$ to $\Omega$.
    Since $I^{\prime}$ and 
$I^{\prime\prime}$ are positive (the moment of inertia
   increases with $\Omega$ and the centrifugal force grows as the
   equatorial radius)
     the  braking index is always less than  the index $n$ of the energy loss
mechanism (\ref{energyloss}). 
And the deficit is independent of $n$.
  
  The important question before us now is whether the departure of the
  braking index $n(\Omega)$ from the intrinsic
  index $n$  is substantial and for what angular velocity range.
  How sensitive is $n(\Omega)$ to the internal constitution of stars?
  In all cases that we have looked at, the braking index  for
  pure magnetic dipole radiation
  has a value near three for low frequencies
  like the Crab, but it falls to values less than unity near the Kepler
  frequency (as a consequence
    of the universal fact that for a rotating star
      the moment of inertia is an increasing function
	of frequency).
  At about half Kepler (the approximate frequency of the
  two 1.6 ms pulsars) the braking index is about two.
  However this does not tell the whole story.
                                                     
  At the frequency of
  millisecond pulsars, $I^\prime$ has a value of about $0.03$ 
  km$^3$\,s. Therefore the value of the dimensionless parameter $\xi\equiv
  I^\prime \omega/(2I) \approx 0.4$ for millisecond pulsars as can be
  read
  from Fig.\ \ref{io}
  (at $\Omega\sim 4000$~rad~s$^{-1}$).
  The rate of change of a millisecond pulsar's angular velocity is therefore
  about 70\% less for the same $n$ and $K$ as given by
  (\ref{braking2}) compared to (\ref{braking}).
  And the magnetic field $B$ as given by $\sim (P \dot{P})^{1/2}$
  is about 20\%   larger   (larger by $(1+\xi)^{1/2}$).
   than the value estimated from (\ref{braking}).
   The usual dipole age formula is not at all valid for a millisecond pulsar,
   since the slope of $I$ differs so strongly from zero for such pulsars.
   One cannot analytically integrate (\ref{braking2}) to get a value
   of the age. The integral of $\dot{\Omega}$ depends implicitly on the
   structure of the star which is changing over time.
  
  \subsubsection{Effect of Continuous Structural Changes}
  Disregarding the feature at $\Omega \sim 1250$ rad/s 
  until the next section,
  the general trend in the moment of inertia is shown in Fig.\
  \ref{io}. 
  As a pulsar spins down the density at every radial distance increases
  and the star's equatorial radius shrinks. Baryon and quark thresholds
  (see  Fig.\ \ref{pops_240}
  and   Fig.\ \ref{pops_k240b180}) as well as  boundaries between
  confined, mixed and deconfined phases and geometrical phases
  (see Fig.\ \ref{cry1})
   occur at unique densities (for given model of the \eosp).
   The boundaries will shift 
  in their radial location as a function of frequency. All such features will
  have their effect on the moment of inertia of a given baryon mass star,
  and hence on the  braking index as shown
  in Fig.\ \ref{no}. (We refer here only to the smooth behavior
  interpolated through the sharp discontinuous behavior at $\Omega
  \sim 1250 {\rm~rad/s}$.) However these relatively continuous changes
  do not account for the departure from three of the apparent index of
  the Crab pulsar and 
the other three pulsars for which the index has been measured.
  I presume that these  
  departures reflect on the energy loss mechanism itself.
  I do not discuss the energy loss mechanism here but only the effects  of
  the changing moment of inertia
  on the corresponding braking index.
  \begin{figure}[tbh]
  \begin{center}
  \leavevmode
  \centerline{ \hbox{
  \psfig{figure=ps.lipari14,width=2.38in,height=2.76in}
  \hspace{.2in}
  \psfig{figure=ps.lipari15,width=2.38in,height=2.76in}
  }}
  \begin{flushright}
  \parbox[t]{2.25in} { \caption { \label{io} Moment of inertia
  as a function of rotational angular
velocity. At angular velocities below $\sim 1250
  {~\rm rad s^{-1}}$ a pure quark phase
  of increasing radius with decreasing frequency (central density)
  occupies the central region of the star.
  }} \ \hspace{.2in} \
  \parbox[t]{2.25in} { \caption { \label{no} The 
  braking index as a function of rotational angular 
velocity. The sharp change at
  $\sim 1250{~\rm rad s^{-1}}$ occurs, as with decreasing angular velocity,
 the core
  begins to dissolve into quark matter 
  (see Fig.\ \protect\ref{OKPOEQ_1_42_G3_B18}.
  }}
  \end{flushright}
  \end{center}
  \end{figure}
  
  We summarize the long-term
  evolution of the pulsar braking index and its dependence on the internal
  constitution of stars. In all cases, the
  braking index will be much less than the
  intrinsic index of the energy-loss mechanism (three
  in the case of magnetic dipole radiation) for  pulsars with short periods,
  especially millisecond pulsars.
  For periods
  of twice Kepler or more, the braking index is close to one unit less than 
  the intrinsic index and increases
  toward the intrinsic index  for very long periods (i.e., most pulsars).
  However, over a short era ($\sim 100,000$ y) 
  the braking  index may be extremely anomalous and exhibit values that
  lie anywhere from negative to positive values  as a result of
  the onset of a phase transition and its growth in radial extent.
  We discuss this next.

  \subsubsection{Braking Index and the Deconfinement Phase Transition}
  How can abrupt large-scale changes in the moment of inertia occur?
  With decreasing rotational frequency the central density of the star
  increases, and since the density and pressure profiles of neutron stars
  are very flat in the central region, the radial point at which a given
  density occurs changes by a considerable fraction of the radius of the
  star for a very small fractional change 
  in its rotational frequency.
  If, during the pulsar spin-down, 
  the central density passes from below to above
  the density for the phase transition, the central region occupied initially
by relatively stiff nuclear matter will be replaced by  more
compressible,  and therefore denser,
quark matter. 
The region
 of quark matter will expand
greatly with small decreases in angular velocity 
because of the flat density profile.
The anomalous concentration of mass in the stellar interior occasioned by the
phase transition 
will be mirrored in structural changes in the
star such as its size and moment of inertia:
the phase transition has ushered in an era in which 
the  star
shrinks anomalously as it spins down over time and its
mass
becomes ever more concentrated near its center---more
so than would be the case for a  star  composed of a simple fluid on which
a weakening centrifugal force was acting. The concentration arising from
the greater compressibility of quark matter is {\sl amplified}
by its greater gravitational attraction  on the
outer parts of the star.

At the stage described, the tendency of the star to shrink as
the  region occupied by  quark matter grows in radius, counteracts by angular
momentum conservation,  the  
deceleration 
 $\dot{\Omega}$ caused by
radiation.  The  growth  of a central
region of deconfined quark matter acts, so to speak,
as a governor in the mechanical sense, in resisting pulsar spin-down.
The  structural changes accompanying
the phase change
thus prolong the epoch over which quark matter
engulfs 
the central region---a situation that is highly
favorable for observation of a signal of the transition epoch.

The behavior of the moment of inertia in the frequency range in which the
phase transition boundary moves outward
in the star is 
shown in Fig.\ \ref{oi}. The temporal development is from large to small
moment of inertia. We see that the star actually enters an era in which
it spin up for a time! This is  analogous to the situation
observed in the rotational spectra of some nuclei.
In nuclei it is a phase transition
from a normal Fermi gas
at high spin to a pair-correlated phase at lower spin that causes
a change in the moment of inertia due to a weakening
Coreolis interaction.

  \begin{figure}[tbh]
  \begin{center}
  \leavevmode
  \centerline{ \hbox{
  \psfig{figure=ps.lipari16,width=2.38in,height=2.76in}
  \hspace{.2in}
  \psfig{figure=ps.lipari17,width=2.38in,height=2.76in}
  }}
  \begin{flushright}
  \parbox[t]{2.25in} { \caption { \label{oi} Detail of the
  moment of inertia
  as a function of rotational angular
velocity in the region of backbending (pulsar spin-up).
  }} \ \hspace{.2in} \
  \parbox[t]{2.25in} { \caption { \label{ni} Detail of the 
  braking index as a function of $I$ (single-valued function of time)
in the spin-up region.
  }}
  \end{flushright}
  \end{center}
  \end{figure}

The particular way in which the  deceleration (\ref{braking2})
and braking index (\ref{index}) are effected by the backbending
of the moment of inertia (Fig.\ \ref{oi}) can be understood with
reference to the formulae. In particular, when $I^\prime$ is large
and negative (the backbend in $I$) the deceleration changes sign---the pulsar
spins up. When $I^{\prime\prime}$, which is related to $I^\prime$ by
\beqn
-I^{\prime\prime}=  I^{\prime~3}
\frac{d^2\Omega}{dI^2}
\eeqn
changes from positive to negative
infinity  at both turning points seen in Fig.\ \ref{oi}, 
the braking index swings from nearly 3
to infinity, to negative infinity and back to 
nearly 3 (Fig.\ \ref{ni}).
This is a remarkable signal considering that the braking index is usually
thought of as a constant (3 for magnetic dipole radiation).

We emphasize that our calculation does not imply
a prediction of the frequency or stellar mass at which the phase
transition will occur. Nor does it even predict that  a phase
transition will occur at all. The results
discussed above pertain to a particular model
star. Very little is known about the high-density
\eosp. Rather our results show  what the signal might be if the
transition does occur
in neutron stars because of the asymptotic freedom of quarks.

We  estimate  the plausibility of observing  
  phase transitions in the  pulsar population.
 The duration over which the observable index is
 anomalous is $\Delta T \approx -\Delta \Omega/\dot{\Omega}$ where
 $\Delta \Omega$ is the angular velocity
 interval of the anomaly ($\approx 100$ rad/s).
 For  a typical period derivative,
 $\dot{P}\sim 10^{-16}$, we find $\Delta T \sim 10^5$ years.
 During  a typical pulsar's active lifetime, about $10^7$ yr, the signal
 (small or negative index) would endure for 1/100 of the lifetime.
 Given that $\sim 10^3$ pulsars are known
 about 10 of them may be signaling the phase transition.

  \section{Strange Stars}
  
  \subsection{The Strange-Matter Hypothesis}
  
  We are so accustomed to the confined phase of hadronic matter (quarks
  confined in hadrons) that we usually do not question whether it is in fact
  the absolute ground state of the strong interaction.
  When we look to the furthest reaches of the Universe
  we see spectral lines that can be identified with molecules, atoms and
  nuclei with which we are familiar.
  Does this not directly inform us that the confined phase is indeed
  the ground state? The answer in a word is ``no''.
  After all, the
  absolute ground state of confined hadronic matter is $^{56}$Fe. Its
  binding energy per nucleon is 931 MeV, lower by some eight MeV than the
  nucleon mass. Yet there is very little iron in the Universe and we know
  perfectly why this is so. It takes a stellar lifetime to convert a very
  small fraction of  the primordial hydrogen  to iron.
  So the contents of the Universe have little bearing on the question.
  
  For different reasons than this Bodmer (1971) and Witten (1984)
  hypothesized independently that the absolute ground state is strange quark
  matter, an approximately equal mixture of the three light flavor quarks,
  u, d and s
  \cite{bodmer71:a,witten84:a}. 
  One can grasp the distinct possibility that this state of matter
  lies quite close in energy per baryon  to  
  iron---the ground state of the confined phase.
  From a variety of nuclear data we are convinced that nuclei
  are composed of neutrons and protons. Since nuclei are two-flavor
  objects, this informs us that {\sl two}-flavor
  quark matter lies higher in energy per baryon number than  the nucleon
  mass. The available energy scale by which they differ is $\Lambda_{{\rm
  QCD}}\sim 100$ to $200$ MeV. On the other hand, for the same reason that
  dense nuclear matter will distribute baryon number over as many baryon species
  as are energetically available, three-flavor quark matter will lie
  lower in energy per baryon than two-flavor quark matter at high density. 
  A Fermi
  gas estimate of the energy difference is again of the order of
  $\Lambda_{{\rm QCD}}$
  \cite{glen90:a}. This places the energy per nucleon of iron and
  that of
  dense strange quark matter at about the same value. Lattice QCD is quite
  unable to predict energies that could be as close as a few percent
  different. 
  The question can only be answered by observation, and I will argue that
  pulsars are the most likely sources of the answer.
  I will discuss limits on rotation of strange stars and neutron stars
  in this connection.
  
  Needless to say, the question concerning the nature of the true ground state
  is a fundamental one.
  As we shall see,  there is no sound basis on which either  to 
  reject or confirm
  the hypothesis at the present. The Universe may indeed occupy a metastable
  though long-lived state.

  \subsection{Viability of the Strange-Matter Hypothesis}
  
  \subsubsection{Stability of Nuclei to Decay to   Strange Nuggets}
  
  One might object that the energy argument given above is not
  satisfactory, since excited states decay to the ground state and we know that
  the confined state exists. Should a nucleus
  not then decay forthwith to  a strange nugget\footnote{Strange objects
  with a number ($ 3A,~A < 10^{50}$) of quarks 
  small enough   that gravity is irrelevant
  are called nuggets -- otherwise stars.} of the same number of quarks
  but distributed over the three light flavors in an approximately
  equal proportion
  if strange matter were the true ground state?
  It  could not do so except on a time scale of the order larger
  than  the
  age of the Universe. A nucleus of A nucleons would have to
  undergo A {\sl simultaneous} weak interactions.
  One interaction at a time would not do  since we know that the first would
  produce a hypernucleus and the $\Lambda$ is more massive than the
  nucleon. 
  
  Of course, nuclei of small A would appear to be candidates for decay
  by the weak interaction
  to strange nuggets, yet they do not. This is easily understood.
  Finite size  effects, especially
  the surface energy, places small-A nuggets  at a higher
  energy per baryon than large-A nuggets.
  Calculations based on the bag model of confinement, though not reliable
  quantitatively, show that the energy per baryon of strange matter
  is a decreasing function of baryon number. It is higher than the
  energy of low-A nuclei, decreasing  from the Lambda
  mass at A=1, and approaching an assymptote for large A that lies
  below the energy per baryon of iron \cite{farhi84:a,madsen94:a}.
  So whether  or not
  strange nuggets of large A or strange matter in bulk is 
  absolutely stable, nuggets of small A are not.

  The density of the strange-matter objects 
  is higher than nuclear density
  so that the Fermi energy of quarks
  is larger that the strange-quark mass ($\sim 150$ MeV).  Therefore, all
  three flavors are about equally populated. Whether the
  critical A  for which strange
  nuggets have lower energy than the corresponding
  nucleus is 100 or a 1000, we cannot be sure from bag model calculations.
  We conclude that low-mass nuclei are protected from decay to strange matter
  because they have a lower energy  than nuggets of the
  same A. High-mass nuclei are 
  protected because a  large number of {\sl simultaneous}
  weak interactions (an A'th order weak interaction) would be required to
  convert them to strange nuggets.

  \subsubsection{The Universe and its Evolution}
  
  Since the matter of the
  very early Universe occupied the deconfined phase, 
  why did  matter not remain in this phase? The answer in a word is that
  the Universe was very ``hot''. As the Universe expanded, opening voids
  in the hot dense quark matter, whose cold dense ground state is, by
  hypothesis the {\sl absolute} ground state, the 
  hot quark matter evaporated into
  nucleons. There remains some debate as to whether large enough
  objects of strange matter could have cooled before evaporation.
  But there is general consensus that little if any primordial strange
  matter survives \cite{alcock85:a,vinhmau94:a}.
  
  For the above reason, the Universe would have evolved along the path of
  confined quark matter which we see today, and of which we are made.
  Strange matter, if the hypothesis is true, can be recreated only on the
  order of stellar lifetimes. Neutron stars, if dense enough in their
  cores, will dissolve into two-flavor quark matter. Because of the high
  density, the Fermi energy of two-flavor quark matter will exceed the
  mass of the strange quark, so that two-flavor quark matter
  will rapidly 
  weak decay
  into strange quark matter, one strangeness-changing interaction at a time
  until the matter is approximately an equal mixture of the three light 
  quark flavors. Being the ground state, and there being no barrier to
  conversion of dense nuclear matter to strange matter, the whole star
  will be consumed
  \cite{olinto87:a}. It will become a strange star. This is in contrast to
  the situation discussed above for light nuclei, where single 
  strangeness-changing reactions are endothermic, the Fermi energy in a nucleus
  being small compared to the strange quark mass.
  
  \subsubsection{Stability of Nuclei to Conversion by Cosmic Strange Nuggets} 
  
  Binary compact stars exist and their orbits decay by gravitational radiation
  \cite{taylor89:a}.
  If at least one of the pair is a strange star, the final collision will 
  likely spew forth some strange matter as fragments into the Universe since 
  collisions of neutron stars are expected to do so \cite{clark77:a}.
  Fragmentation usually will produce a preponderance of small fragments.
  These could  have anywhere from the minimum baryon number for which strange
  nuggets
  are stable to more massive fragments. Therefore, there would exist
  a strange nugget component in cosmic
  rays. Some nuggets would impinge on other stars and in particular
  on the Earth. (An estimate of the flux can be found
  in \cite{glen90:a}.)
  What prevents their consuming the hadronic matter with
  which they come into contact? 
  
  Because  the
  strange quark has greater mass than the other two light flavors,
  quark matter has a slight deficit of strange quarks compared to the
  others. 
  Consequently, strange nuggets carry positive charge.
  Strange nuggets and nuclei therefore repel each other.
  I have estimated  the concentration of low-mass strange nuggets
  in the Earth's surface layer, taking account of geological mixing to
  ten kilometer depths, and find that  strange nuggets
  would be very rare objects in 
  earthly samples, much less than $10^{-15}$ nuggets per nucleon.
  The moon has been exposed as long as the earth to these cosmic nuggets,
  and its surface has been tranquil for most of its life. Moon rock is a
  more promising source.
  
  \subsection{Limits to Neutron Star Rotation}
  
  \subsubsection{Absolute Limits}
  
  A neutron star at the mass limit can rotate most rapidly of all
  stars in its sequence since it has the smallest radius. (See
  Fig.\ \ref{radnq}). This can be seen from the condition that gravity balances
  the
  centrifugal force at the equator.
  The Kepler angular velocity can be approximated
  by \cite{friedman89:a,haensel89:a,weber90:d},
  \beqn
  \Omega_K\approx \zeta (M/R^3)^{1/2}\,.
  \label{clcandg}\label{Kepler}
  \eeqn
  where $\zeta \approx 0.625$.
  Although this result is  Newtonian 
  reduced by the prefactor,
  it agrees with numerical calculations
  in   General Relativity 
  to  better than 10\% \cite{haensel89:a}.
  We are here interested in establishing
  the minimum rotational period for a star
  that is gravitationally bound,  by a variational
  calculation made under conservative physical
  assumptions \cite{glen92:a}.
  We adopt the following minimal constraints:
  \begin{enumerate}
  \item Einstein's general-relativistic
  equations for stellar structure hold.
  \item
  The matter of the star
  satisfies $dp/d\epsilon \geq 0$
  which is a necessary condition
  that a body  is  stable
  both as a whole,  and also with respect to the spontaneous
  expansion or contraction of elementary regions away from equilibrium
  (Le Chatelier's principle).
  \item  The \eos  satisfies the
  causal constraint for a  perfect fluid;
  a sound signal cannot propagate
  faster than the speed of light,
  $v(\epsilon)
  \equiv\sqrt{dp/d\epsilon} \leq 1 $,
  which is also the appropriate expression for sound signals
   in General
  Relativity \cite{curtis50:a}.
  \item The high-density
  \eosp,  whatever it is, matches continuously
  in energy and pressure to the
  low-density equation of state of Baym, Pethick and Sutherland \cite{baym71:b}.
  \end{enumerate}
  \index{Causality}
  
  The results of the variational search for the minimum period of neutron stars
  as a function of their mass is shown in Fig.\ \ref{periodp}. A canonical
  neutron star can have a period not less than 0.3 ms. The actual physical limit
  is likely to be higher than this because of gravitational wave instabilities.
  So our result is a most conservative one. We note from Fig.\ \ref{radnq}
  the fine tuning problem required to achieve the limit. The mass window
  for rapid rotation is
  extremely narrow.
  \begin{figure}[tbh]
  \begin{center}
  \leavevmode
  \centerline{ \hbox{
  \psfig{figure=ps.lipari18,width=2.38in,height=2.76in}
  \hspace{.2in}
  \psfig{figure=ps.lipari19,width=2.38in,height=2.76in}
  }}
  \begin{flushright}
  \parbox[t]{2.25in} { \caption { \label{radnq} Generic mass-radius relation
  for neutron stars. Those that lie below the curves marked 1.6 ms and 0.5 ms
  can rotate as fast or faster than the these periods.
  \protect\cite{book}.\protect\courtesy.
  }} \ \hspace{.2in} \
  \parbox[t]{2.25in} { \caption { \label{periodp} Neutron stars can fall only
  in the indicated region. All stars are forbidden in the region so marked.
  Hypothetical stars that are not bound by gravity can fall within the
  blank region  \protect\cite{book}.\protect\courtesy.
  }}
  \end{flushright}
  \end{center}
  \end{figure}
  
  Even if a pulsar of canonical mass
  were observed with a period near the limit, it would be implausible 
  to interpret it as a neutron star. The central density would be about
  20 times nuclear density. (See Fig.\ \ref{cdensity}.)
  Considering the charge radius of the 
  proton (0.8 fm), nucleons would be squeezed out of existence into their
  quark constituents.
  
  \subsubsection{Practical Limits}
  
  The limit above is  obtained under extremely conservative
  conditions. There is no minimum principle for rotation as there is for
  energy (mass). A star does not need to be so configured as to rotate
  fast. So any knowledge additional to that of the four conditions above
  will raise the lower bound on period,  and realistic equations of state
  do so considerably
  \cite{weber90:d}. 
  Realistic \eoss yield stellar models with Kepler
  frequencies in the 1 ms range.  (Those models that have shorter Kepler periods
  have central densities so high as to exceed the applicability of
  models of matter based on nucleons as constituents.)
  
  There is another instability beside the
  mass-shedding instability (Kepler) that places a more severe
  limit on rotation. It is a gravitational wave instability.
  It raises the lower limit on rotational  periods of 
  neutron stars by an additional 30\% or so
  \cite{weber90:c,weber90:a}. However it is difficult to be precise as to
  how much this instability raises the limiting period because of the
  uncertainty of the viscosity of dense stellar material.
  
  We conclude that if a period as small as one millisecond were discovered, it
  would be quite difficult to reconcile
  with realistic models of neutron stars.
  
  \subsection{Limits to Rotation of Strange Stars}
  
  Strange stars are not bound by gravity but by the strong interaction.
  Gravity simply squeezes them, hinders their fission and imposes a limit on 
  their mass, above which they would collapse. 
  Let us discover the constraint that
  rapid rotation places on strange matter \cite{glen90:a}.

  Denote the normal energy density of self-bound matter
  (the density at which the internal pressure
  vanishes) by $\epsilon_b$. A small nugget therefore has mass
  \beqn
  M= \frac{4}{3} \pi R^3 \epsilon_b ~~~~~({\rm no~gravity})
  \label{nogravity}
  \eeqn
  so that, unlike neutron
  stars (or more generally stars bound only by gravity), the 
  mass-radius relation for small mass is
  \beqn R \propto M^{1/3}\,.
  \label{rm}
  \eeqn
  This dependence of the radius on mass
  has
  a  generically different form from that
  of gravitationally bound stars as shown in Fig. \ref{radnq}.
  We can rearrange (\ref{Kepler}) to read
  \cite{glen89:k,glen90:a}:
  \beqn \epsilon_b \geq \frac{3}{4 \pi G}
  \Bigl (\frac{\Omega}{\zeta}\Bigr )^2 \label{stlim}
  = 1.4 \epsilon_0 \Bigl ( \frac{\rm ms}{P} \Bigr)^2. \eeqn
  where $\epsilon_0$ denotes the normal density of nuclear matter
  ($2.5\times 10^{14}~{\rm g/cm^3}$).
  This provides the condition  that must be satisfied
  by  the ``normal'' density of strange matter,
  the density at which it is in the equilibrium ground state,
  so that it can have a designated period $P$.

  \begin{figure}[htb]
  \begin{center}
  \leavevmode
  \centerline{ \hbox{
  \psfig{figure=ps.lipari20,width=2.38in,height=2.76in}
  \hspace{.2in}
  \psfig{figure=ps.lipari21,width=2.38in,height=2.76in}
  }}
  \begin{flushright}
  \parbox[t]{2.25in} { \caption { \label{cdensity} Minimum central density
  of (the non-rotating counterpart) of a neutron star rotating at the
  boundary of Fig.\ \protect\ref{periodp} (From Ref. \protect\cite{glen92:a}).
  }} \ \hspace{.2in} \
  \parbox[t]{2.25in} { \caption { \label{radqh2} Mass-radius
  relationships for neutron stars
  and  strange stars.
  Stars below dotted  lines can rotate at
  or faster than
  the  indicated period. These curves are representations of
  (\protect\ref{Kepler}).
  Equation of state for self-bound matter is  parameterized as
  $\epsilon=p/v^2+\epsilon_b$.
  (From \protect\cite{glen90:a}.)
  }}
  \end{flushright}
  \end{center}
  \end{figure}
  The mass-radius relationship for strange stars (self-bound) is entirely
  different from that of neutron stars  as  is shown
  in Fig.\ \ref{radqh2}. For neutron stars, the smaller  the mass
  the less gravity  compresses the star and the larger it will be.
  However strange stars (by hypothesis) are bound by the strong interaction
  and therefore at some particular value of the energy density. Consequently,
  for small A, their radii scale as in (\ref{rm}).
  Because of the
   mass-radius relationship of strange objects,
   which is modified only
  slightly by gravity for the more massive objects,
  the entire sequence of strange stars can rotate about as fast as that
    at the  
  limiting mass. In contrast, the most massive
  neutron stars can rotate much more rapidly than other members of the
  sequence as seen from (\ref{Kepler}).
  Hence, an observation of a neutron star
  near the limiting value of the frequency would be a rare occurrence.

  \begin{figure}[htb]
  \begin{center}
  \leavevmode
  \centerline{ \hbox{
  \psfig{figure=ps.lipari22,width=2.38in,height=2.76in}
  \hspace{.2in}
  \psfig{figure=ps.lipari23,width=2.38in,height=2.76in}
  }}
  \begin{flushright}
  \parbox[t]{2.25in} { \caption { \label{prof} Three strange star energy
  distributions \protect\cite{book}.\protect\courtesy.
  }} \ \hspace{.2in} \
  \parbox[t]{2.25in} { \caption { \label{profqh} Comparison of a strange and
  neutron star energy distribution \protect\cite{book}.\protect\courtesy.
  }}
  \end{flushright}
  \end{center}
  \end{figure}
  \subsection{Strange-Star Configurations}
  
  Because strange stars are (by hypothesis) bound by the strong interaction,
  and only additionally by gravity,
  they have very sharp surfaces. The surface of a star corresponds to 
  vanishing pressure because  vanishing pressure can support no
  overlaying layer of matter against the gravitational attraction
  from within.
  Therefore the 
  density at the inner edge of the surface is equal to the equilibrium
  density of strange matter\footnote{Equilibrium implies vanishing
  pressure.}---several times that of nuclear matter.
  The transition from a high density to zero at the edge of the star
  occurs 
  in a
  strong interaction distance ($\sim 10^{-13}$ cm) since the strong
  interaction binds the star. At first sight this may seem 
  unusual. But it is quite analogous to nuclei. Both are bound by
  the strong short-range
  force  and it matters not at all to the skin thickness how much
  matter lies behind the surface. Bare strange stars, if they
  exist, have the sharpest surfaces of any conceivable
  object. Ordinary material objects have surface thicknesses
  corresponding to the range of molecular forces.
  Figure \ref{prof}
  shows the density profiles of several strange stars and Fig.\ \ref{profqh}
  compares a neutron star and strange star.
  Quark populations in a strange star are shown in Fig.\ \ref{b1545_a0}.
  
  Charm quarks are too massive compared to the chemical potential to
  be present. It is of some interest to enquire whether any star can contain
  charm quarks. We compare the baryon mass and gravitational mass of
  strange stars past the first limiting mass in Fig.\ \ref{mas_b1545_a0}.
  We have made a stability analysis and find, as expected, that no
  configuration past the first mass limit is stable against acoustics
  modes that would cause its collapse to a black hole
  \cite{glen94:b}. So charm quarks
  have no present astrophysical interest.

  \begin{figure}[tbh]
  \begin{center}
  \leavevmode
  \centerline{ \hbox{
  \psfig{figure=ps.lipari24,width=2.38in,height=2.76in}
  \hspace{.2in}
  \psfig{figure=ps.lipari25,width=2.38in,height=2.76in}
  }}
  \begin{flushright}
  \parbox[t]{2.25in} { \caption { \label{b1545_a0}
  Quark populations in a strange star of mass 1.6~M$_{\odot}$
  \protect\cite{book}.\protect\courtesy.
  }} \ \hspace{.18in} \
  \parbox[t]{2.25in} { \caption { \label{mas_b1545_a0} Strange-star sequence
  showing the baryon and gravitational mass. All stars beyond the first
  maximum are unstable \protect\cite{book}.\protect\courtesy.
  }}
  \end{flushright}
  \end{center}
  \end{figure}
  
  \subsection{Strange Stars with Nuclear Crusts}
  
  A strange  star  has a sharp edge of thickness defined by
   the range of the strong interaction (cf. Fig.\ \ref{prof}). Such `bare'
   strange stars  are unlikely to exist as such. 
   Strange stars can  carry a crust of nuclear
   material which is suspended from contact with the strange
   star by a strong electric dipole field. This was noted
   by Alcock, Farhi
  and Olinto\index{Strange star}
  \cite{alcock86:a}  who
  pointed out that the electrons (which neutralize the
   positive charge  of strange quark matter
   and are bound to it by the Coulomb attraction) extend several hundred
   fermis (the de Broglie wavelength) beyond the edge.
   In consequence, just inside the surface
   there is a positively charged layer (because strange matter
   by itself is slightly positively charged).
    A dipole layer of high
   voltage is thus created. The surface dipole can support  a layer
   of ordinary
   matter (which it polarizes) out of contact with the core.
   The separation gap prevents the conversion of the nuclear surface
   layer to quark matter.
                     
  \begin{figure}[htb]
  \begin{center}
  \leavevmode
  \psfig{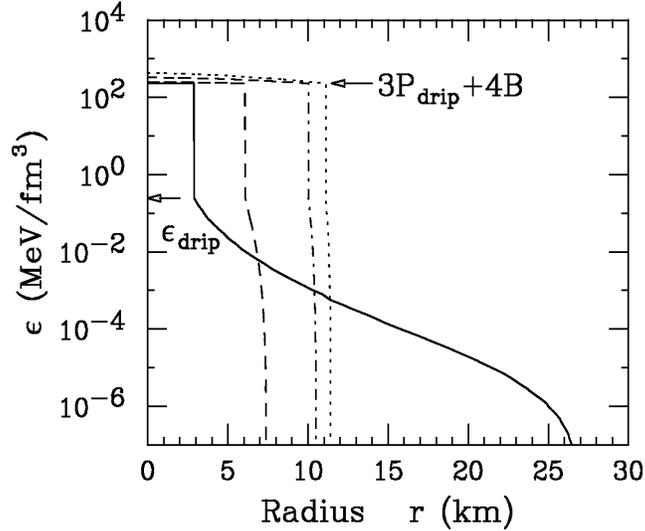}
  \parbox[t]{4.6 in} { \caption { \label{evsr_b145_nn} Strange stars with nuclear crusts at the drip density:
  energy density as a
  function of radial distance from the
  star's center for gravitational masses $M$/M$_{\odot}$=0.020 (solid line),
  0.20 (dashed), 1.00 (dash-dotted), and 1.50 (dotted). The bag constant
  is $\bag=145$ MeV. (From \protect\cite{glen92:b}.)
  }}
  \end{center}
  \end{figure}
  The gap between the core
  and its crust
   of nuclear material is estimated to be of the order of several
   hundred fermis \cite{alcock86:a}.
   Effects of finite temperature have been investigated in
   ref. \cite{glen94:b}. The gap prevents the conversion of the crust to
   strange matter unless the crust density is too high.
   The maximum density of the nuclear  crust
   is strictly
   limited by the neutron drip density  $\edrip \approx 4\times 10^{11}$ \gpercm~
    above which  free neutrons would gravitate
   to the strange core and be converted to quark matter.
   It is likely that strange stars do have such crusts of various inner
   crust densities depending on their histories and ages. Interstellar
   space is not empty.

  Thus, strange stars with nuclear crusts form a two-parameter sequence
  corresponding to the central density and the inner crust density. In
  practice, we fix the inner density of the crust and vary the central
  density to generate a corresponding sequence. For a particular
  equation of state of core and crust material,
  each pair of such
  parameters defines a unique stellar structure with a particular mass
  and radius.
  
  \begin{figure}[htb]
  \begin{center}
  \leavevmode
  \psfig{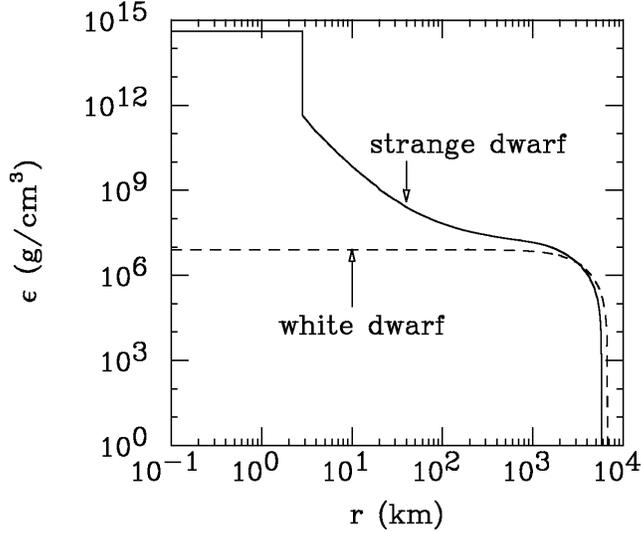}
  \parbox[t]{4.6 in} { \caption { \label{er_100794}  Comparison of mass
  profiles of a strange dwarf (with crust at the neutron 
  drip density) and a white dwarf of mass typical of white
  dwarfs ($M=0.6$~M$_{\odot}$). (From \protect\cite{glen94:f}.)
  }}
  \end{center}
  \end{figure}
  A selection of strange stars with crusts at the limiting density 
  $\edrip$ are shown in Fig.\ \ref{evsr_b145_nn}. These are the counterparts
  to neutron stars. They consist mostly of a massive strange quark core.
  Because of the gravitational attraction of the core on the crust,
  the crust is generally thin but grows in thickness for smaller cores.
  Along a sequence of ever
  smaller cores the mass  and thickness of the crust  can grow to 
  white-dwarf dimensions. What is unusual about such stars is that the 
  density of nuclear matter in the crust can be as high
  as the drip density ($4\times 10^{11}$ \gpercm) which is very much larger
  than the central density of white dwarfs
  ($\epsilon_{{\rm wd}}\leq 10^{9}$ \gpercm). 
  Such stars with strange matter cores
  and layer of dense nuclear material of white dwarf dimensions
  form a different class of dwarfs, called strange dwarfs
  \cite{glen94:c}.
  Were it not for the gravitational attraction of the dense strange quark core,
  they would be unstable. 
  A comparison between a white dwarf and strange dwarf is shown in 
  Fig.\ \ref{er_100794}.
  
\begin{figure}[htb]
  \begin{center}
  \leavevmode
  \psfig{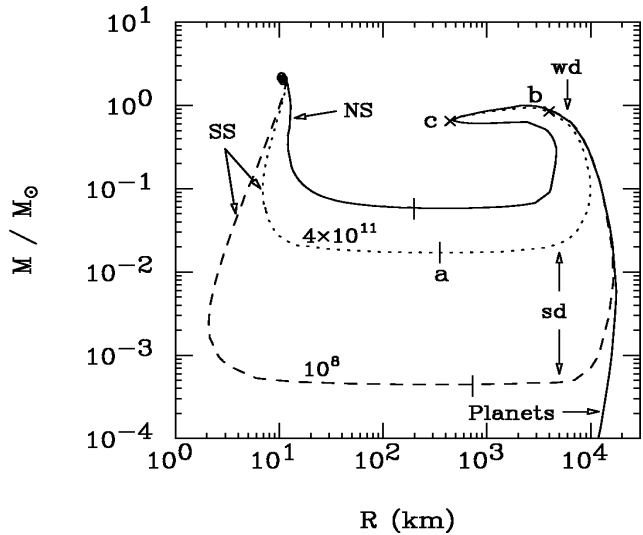}
  \parbox[t]{4.6 in} { \caption { \label{s95_1} Neutron star
  (NS) -- white dwarf  (wd) sequence,
  (solid line).
  Two  strange star (SS) -- strange dwarf (sd) sequences,
  for which the inner crust density of nuclear material  has the indicated
  values (in g/cm$^3$).
  The higher value is the
  drip density.
  Vertical bars mark minimum mass stars. Crosses mark termination of the
  strange star sequences where the strange core shrinks to zero. At those points
  strange dwarfs become identical to ordinary white dwarfs. (From
  \protect\cite{glen94:c})
  }}
  \end{center}
  \end{figure}

  A sequence of neutron stars to white dwarfs (and eventually planets)
  is shown in Fig.\ \ref{s95_1} together with two members
   of a continuum of strange
  objects. The one has  inner crust density fixed at the drip density --
  higher than any white-dwarf density; the other has  inner crust density
  of
  $10^8$~g~cm$^{-3}$, which is the density of a normal white dwarf.
  The latter would be stable without the strange core; the former not.
  What is most remarkable is that the entire sequence of strange objects
  is stable,
  from the maximum mass of the compact strange star to the termination
  of the sequence on the white-dwarf family or at the maximum-mass
  strange dwarf (whichever occurs first in moving from the maximum-mass 
  strange star toward the dwarfs). This was demonstrated
  by a stability analysis
  \cite{glen94:c}. 
  In contrast, the white dwarf -- neutron star sequence has a region of
  stability ranging over many orders of magnitude in central stellar density
  from the minimum mass neutron star to the maximum mass white dwarf.
  
  \subsubsection{Stability of Strange Dwarfs}
  
\begin{figure}[htb]
  \begin{center}
  \leavevmode
  \psfig{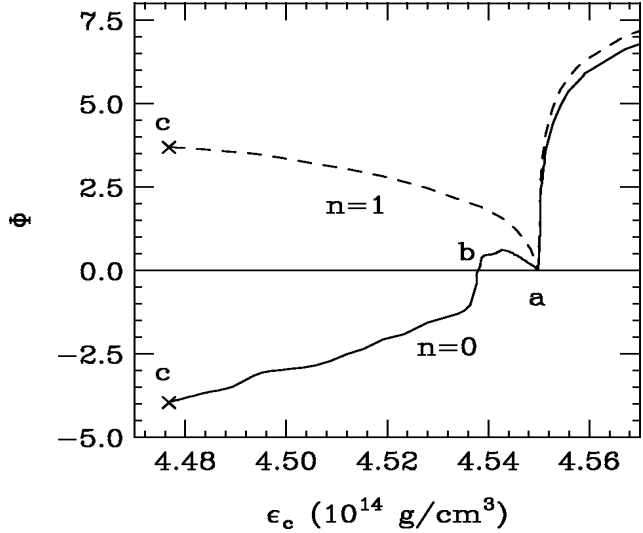}
  \parbox[t]{4.6 in} { \caption { \label{phi} Pulsation 
  frequencies for the two lowest modes $n=0,1$ measured by
  $\Phi(x)\equiv {\rm sgn}(x) \log(1+|x|)$ where $x\equiv (\omega_n/{\rm
  s}^{-1} )^2$ as a function of central star density in the vicinity of strange
  dwarfs having inner crust density equal to neutron drip. For $\Phi<0$,
  the squared frequency is negative and the mode unstable.
  (From \protect\cite{glen94:c}.)
  }}
  \end{center}
  \end{figure}

  The configurations
  of the neutron star-white dwarf sequence that lie between
  the points a to b marked on 
  Fig.\ \ref{s95_1} are unstable.
  We have carried out a stability analysis for the strange sequence
  \cite{glen94:c}. The result for the fundamental (nodeless)
  vibrational mode is shown in Fig.\ \ref{phi}.
  When the function $\Phi$ is positive, so  is the
  angular
velocity $\omega^2$ of the radial vibrational modes. This corresponds to 
  stability. We see that the entire range of strange objects from the
  maximum-mass strange star to the maximum-mass strange dwarf are
  stable.
  
  \subsubsection{Strange Dwarfs as Microlensing Candidates}
  
  Strange dwarfs could be made in several ways \cite{glen94:c}. The capture
  by main-sequence stars of
  strange nuggets as a component of cosmic rays, discussed earlier,
  is one possibility. Main-sequence stars are long-lived, large-area 
  collectors of a cosmic flux of nuggets. Once captured a nugget would 
  gravitate to the center of the star and rest dormant for almost or all
  of the stellar life time. If the star has mass greater than
  $\sim 8$~M$_{\odot}$, the core will contain free neutrons in the last few hours
  of the existence of the star. Upon core collapse a strange star will
  be born. If the progenitor mass is smaller
  than $\sim 8 \msun$, nuclear burning is incomplete
  and free neutrons will not be present. A white dwarf is born as a result
  of vibrational instabilities which expel most of the star in a 
  planetary nebula. The white dwarf is a strange dwarf, having a strange
  core of baryon number or mass corresponding to the lifetime
  acquisition  of strange nuggets by the main-sequence star.
  The number of such nuggets will tend to fill the entire space 
  between the dashed and dotted curves of Fig.\ \ref{s95_1}. They represent
  an enormous `phase space'  of  
  low-mass   objects---Jupiter mass to several hundredths of a solar mass---that
  are candidates for microlensing detection.  
  
  {\bf Acknowledgements}:
         \doe
  

  \end{document}